\documentclass[twocolumn,fp]{jpsj3}

\usepackage{graphicx}
\usepackage{dcolumn}
\usepackage{bm}
\usepackage{color}
\usepackage{ulem}
\usepackage{txfonts}
\def\vector#1{{\boldsymbol{#1}}}
\def\va{{\vector a}}
\def\vb{{\vector b}}
\def\vc{{\vector c}}

\def\vA{{\vector A}}

\def\vH{{\vector H}}
\def\vk{{\vector k}}

\def\vq{{\vector q}}

\def\vr{{\vector r}}
\def\vR{{\vector R}}

\def\vv{{\vector v}}

\def\qbar{{\bar q}}

\def\dps{\displaystyle}

\def\kB{{k_{\rm B}}}

\def\eq.#1{Eq.~(\ref{#1})}
\def\eqs.#1{Eqs.~(\ref{#1})}

\def\refeq#1{(\ref{#1})}

\def\Hc2{{H_{\rm c2}}}
\def\difHc2{{H'_{\rm c2}}}

\def\*ref*{{\color{red}$\leftarrow$Ref.{\bf [~~~~~]}}}

\newcommand\Equation[2]{
\begin{equation}\label{#1} 
#2
\end{equation}
}

\newcommand\Equationnoeqn[2]{
\begin{equation*}\label{#1} 
#2
\end{equation*}
}

\title{
Fulde--Ferrell--Larkin--Ovchinnikov State in Perpendicular 
Magnetic Fields \\ 
in Strongly Pauli-Limited Quasi-Two-Dimensional Superconductors \\ 
}

\author{Hiroshi Shimahara}

\inst{
Graduate School of Advanced Science and Engineering, Hiroshima University, \\ 
Higashi-Hiroshima 739-8530, Japan
}

\recdate{November 5, 2020}

\abst{
We examine 
the Fermi-surface effect called the nesting effect 
for the Fulde--Ferrell--Larkin--Ovchinnikov (FFLO) state 
in strongly Pauli-limited 
quasi-two-dimensional superconductors, 
focusing on the effect of three-dimensional factors, 
such as interlayer electron transfer, 
interlayer pairing, 
and off-plane magnetic fields 
including those perpendicular to the most conductive layers 
(hereinafter called the perpendicular fields). 
We examine the systems with a large Maki parameter 
so that the orbital pair-breaking effect is negligible, 
except for the locking of the direction of the FFLO vector $\vq$ 
in the magnetic-field direction.
It is known that the nesting effect for the FFLO state 
can be strong in quasi-low-dimensional systems 
in which the orbital pair-breaking effect is suppressed 
by applying the magnetic field parallel to the layers. 
Hence, it has sometimes been suggested that 
the nesting effect may hardly enhance the stability of the FFLO state
for perpendicular fields. 
We illustrate that, contrary to this view, 
the nesting effect can strongly stabilize the FFLO state 
for perpendicular fields as well as for parallel fields 
when $t_z$ is small 
so that the Fermi surfaces are open in the $k_z$-direction, 
where $t_z$ denotes the interlayer transfer energy. 
In particular, the nesting effect in perpendicular fields 
can be strong in interlayer states. 
For example, 
in systems with cylindrical Fermi surfaces 
warped by $t_z \ne 0$, 
interlayer states with $\Delta_{\vk} \propto \sin k_z$ 
exhibit $\mu_{\rm e} H_{\rm c} \approx 1.65 \Delta_{\alpha 0}$ 
for perpendicular fields, 
which is much larger than 
typical values for parallel fields, 
such as 
$\mu_{\rm e} H_{\rm c} = \Delta_{{\rm s}0}$ of the s-wave state 
and 
$\mu_{\rm e} H_{\rm c} \approx 1.28 \Delta_{\rm d0}$ of the d-wave state 
in cylindrical systems with $t_z = 0$. 
Here, $\mu_{\rm e}$ and $H_{\rm c}$ 
are the electron magnetic moment 
and upper critical field of the FFLO state at $T = 0$, 
respectively, 
and 
$\Delta_{\alpha 0} \equiv 2 \omega_{\rm c} e^{- 1/\lambda_\alpha}$. 
We discuss the possible relevance of the nesting effect 
to the high-field superconducting phases 
in perpendicular fields observed in the compounds 
${\rm CeCoIn}_5$ and ${\rm FeSe}$, 
which are candidates for the FFLO state. 
The present result could potentially provide a physical reason 
for the fact that the areas in the phase diagrams 
occupied by the high-field phases 
for the perpendicular and parallel fields 
are of the same order.
}

\begin{document}
\sloppy
\maketitle

\section{Introduction}\label{sec:Introduction}

The Fulde--Ferrell--Larkin--Ovchinnikov (FFLO) 
state~\cite{Ful64,Lar64} 
has been examined in quasi-two-dimensional systems 
for candidate compounds, 
such as some organic and 
heavy fermion superconductors~\cite{Cas04,Mat07,Leb08,Wos18,
Sin00,Rad03,Bia03,Yon08a,Kas20}. 
In many cases in theoretical studies, 
two-dimensional models~\cite{Note2D} 
are useful as effective models 
of quasi-two-dimensional systems; 
however, 
depending on the problem under examination, 
the three-dimensionality of the systems due to 
interlayer electron transfer must be treated explicitly, 
for example, 
when interlayer pairing and/or off-plane magnetic fields 
play essential roles.

Since the FFLO state is a superconducting state 
induced by Cooper pairs with 
a finite center-of-mass momentum $\vq$~\cite{Notemultq} 
(hereinafter called the FFLO vector), 
it significantly depends on the Fermi-surface structure, 
for example, because of the direction of $\vq$ relative to 
the anisotropic Fermi surface. 
At the same time, 
unless the orbital pair-breaking effect is 
too weak or too strong~\cite{Gru66,Shi09,NoteqpH}, 
the direction of $\vq$ is locked in the direction of 
the magnetic field $\vH$; 
i.e., $\vq \parallel \vH$.~\cite{Gru66,NoteLO} 
From this point of view, 
it is physically interesting 
that all the candidates discovered thus far are 
quasi-low-dimensional, 
which can be attributed to 
the following two reasons: 
(i) the suppression of the orbital pair-breaking effect 
when $\vH$ is parallel to 
the most conductive layers~\cite{Leb08,Shi97b} 
(hereinafter called the ab-plane) 
and 
(ii) the Fermi-surface nesting effect for the FFLO 
state~\cite{NoteNesting1,NoteNesting2,Shi94}.

For two of the strongest candidate compounds, 
${\rm CeCoIn}_5$ and ${\rm FeSe}$, 
reason (i) would not apply, 
because the conduction electrons have 
large effective masses~\cite{Rad03,Bia03,Kas14,Han18,Shib20}, 
which result in large Maki parameters; 
hence, $\vH$ would not need to be parallel to the ab-plane 
for the emergence of the FFLO state. 
In fact, 
high-field superconducting phases, 
which can be considered to be the FFLO state, 
were observed~\cite{Kum06,Kas14} 
in these compounds 
when $\vH \parallel \vc$ 
as well as when $\vH \perp \vc$, 
where $\va$, $\vb$, and $\vc$ denote the lattice 
vectors in the directions of 
the crystal a-, b-, and c-axes.

The Fermi-surface nesting effect for 
the FFLO state mentioned in reason (ii) 
is an effect analogous with the nesting effect for 
spin- and charge-density waves (SDW and CDW). 
In quasi-low-dimensional systems, 
when nesting instabilities such as 
SDW and CDW instabilities are suppressed 
by sufficient distortion of the Fermi surfaces~\cite{Shi94}, 
the highly anisotropic Fermi-surface structures 
help stabilize the FFLO state. 
The nesting effect for the FFLO state 
can be examined by considering 
the overlap of one of the Fermi surfaces 
of the up- and down-spin electrons 
and another that is shifted by $\vq$.~\cite{Shi94,NoteNesting2} 
In a simple two-dimensional model 
in which the interlayer electron transfer is neglected, 
the Fermi surfaces can touch on a vertical line 
for $\vH \perp \vc$ 
because $\vq \parallel \vH$ as mentioned above, 
whereas for $\vH \parallel \vc$, they cannot touch each other. 
Hence, 
it may be thought that the nesting effect does not work 
when $\vH \parallel \vc$ in quasi-two-dimensional systems. 
For example, 
in ${\rm CeCoIn}_5$ and ${\rm FeSe}$, 
the high-field phases for $\vH \parallel \vc$ 
may appear to be inconsistent with 
reason (ii) as well as with reason (i). 
However, as shown in the following, 
the nesting effect can work even when $\vH \parallel \vc$, 
in the presence of the warp of the Fermi surfaces 
in the direction of $\vc$.

Motivated by these experimental and theoretical studies, 
we examine the effects of three-dimensional factors, 
such as 
interlayer electron transfer, 
interlayer pairing, 
and off-plane magnetic fields including perpendicular fields. 
Off-plane magnetic fields can cause extreme phenomena, 
such as vortex states with higher Landau-level indices, 
through the orbital pair-breaking 
effect~\cite{Shi97b,Kle00,Kle04,Shi09}. 
However, because the orbital effect has been examined 
in previous studies, 
in the present study 
we focus on the nesting effect in the absence of 
the orbital effect. 
We also incorporate the possibility of interlayer pairing 
because, owing to the finite $\vq$, 
strong interplay between the three-dimensional structures 
of the gap function and the Fermi surface is expected. 
In addition, the FFLO state is worth studying for interlayer pairing 
because of the compound ${\rm Sr_2RuO_4}$, 
for which both the FFLO state~\cite{NoteSRO} 
and interlayer pairing~\cite{Has00,Shi20c} 
can be considered.

For quasi-two-dimensional systems in perpendicular fields, 
Song and Koshelev proposed a theory of interplay between 
orbital-quantization effects and the FFLO state, 
and discussed the FFLO state in ${\rm FeSe}$ 
when $\vH \parallel \vc$~\cite{Son18}. 
In the present study, 
we assume weaker magnetic fields.

In Sect.~\ref{sec:formulation}, 
we briefly review formulas and define 
the systems and states to be examined. 
In Sect.~\ref{sec:tz0}, 
we examine systems with Fermi surfaces straight in the $k_z$-direction. 
In Sect.~\ref{sec:finitetz}. 
we examine systems with warped Fermi surfaces. 
We discuss the possible relevance of our results 
to the high-field phases 
in the compounds ${\rm CeCoIn}_5$ and ${\rm FeSe}$. 
The final section summarizes and concludes the paper. 
We define the $x$-, $y$-, and $z$-axes along 
the crystal a-, b-, and c-axes. 
The lattice constants $a$, $b$, and $c$ are absorbed into 
the definitions of the momentum components 
$k_x$, $k_y$, and $k_z$. 
We use units in which $\hbar = \kB = 1$, 
and we denote the electron magnetic moment 
by $\mu_{\rm e} = g \mu_{\rm B}/2$. 
For convenience, 
we define the functions 
\Equation{eq:fn}
{
     f_n(p) 
       = - \int_0^{\pi} \frac{d x}{\pi} 
           \cos (nx) 
           \ln \bigl | 1 - p \cos x \, \bigr | . 
     }
Their explicit forms are 
shown in Appendix\ref{app:Notefn}.

\section{Formulas and Model}\label{sec:formulation}

{\it Formula for the Critical Field} --- 
We use the formula~\cite{Shi97a} 
for the upper critical field at $T = 0$ 
\Equation{eq:hcFFLO}
{
     h_{\rm c} 
     \equiv \mu_{\rm e} H_{\rm c} 
     = \frac{1}{2} \Delta_{\alpha 0} 
         \max_{q} \bigl [ e^{f_{\alpha}(\vq)} \bigr ]  
     }
with 
\Equation{eq:falphageneral}
{
    f_{\alpha}(\vq) = 
    - \frac{1}{\langle |\gamma_{\alpha}({\hat \vk})|^2 \rangle} 
      \left \langle 
        |\gamma_{\alpha}({\hat \vk})|^2 
        \ln \Bigl [ 
        1 - \frac{\vv_{\rm F} \cdot \vq}{2 h_{\rm c}}
            \Bigr ] 
       \right \rangle 
}
for the $\alpha$-wave state with 
the gap function 
$\Delta_{\vk} = \Delta_{\alpha} \gamma_{\alpha}({\hat \vk})$, 
where 
$\gamma_{\alpha}$ denotes a basis function of 
${\hat \vk} \equiv \vk/|\vk|$, 
and 
$\Delta_{\alpha 0} \equiv 2 \omega_{\rm c} e^{-1/\lambda_{\alpha}}$ 
is a scale of the gap function~\cite{Notelambda}. 
The average is defined as 
\Equation{eq:avedef}
{
     \langle g(\hat \vk) \rangle 
     = 
        \int \frac{d^2 {\hat \vk}}{S_0} 
         \frac{\rho(0,{\hat \vk})}
              {N(0)} 
              g({\hat \vk}) 
     }
for an arbitrary $g({\hat \vk})$, 
where 
$\rho(\xi,{\hat \vk})$ is the angle-dependent density of states, 
$N(\xi)$ is the density of states, 
and 
$S_0$ is an appropriate normalization constant.~\cite{NoteS0} 
The FFLO vector is the vector $\vq$ that gives 
the highest $h_{\rm c}$ 
in accordance with the variational principle~\cite{Shi94}; 
however, 
only the magnitude $q \equiv |\vq|$ is optimized in \eq.{eq:hcFFLO} 
because $\vq \parallel \vH$ in the present problem. 
The formula in \eq.{eq:hcFFLO} 
is derived in the weak coupling theory~\cite{Shi97a,Shi99aT0}, 
where a second-order transition is assumed. 
If the second-order transition at $h_{\rm c}$ is to occur, 
$h_{\rm c}$ must exceed 
the Pauli limit $h_{\rm P} = \mu_{\rm e} H_{\rm P}$~\cite{Notechi}.

The Pauli paramagnetic limit $h_{\rm P}$ in anisotropic superconductors 
is given by the formula~\cite{Shi97a,Shi99aT0}
\Equationnoeqn{eq:HP}
{
     h_{\rm P} = \frac{\sqrt{\langle |\gamma_{\alpha}|^2 \rangle}} 
                      {{\bar \gamma_{\alpha}}} 
                 \frac{\Delta_{\alpha 0}}{\sqrt{2}} , 
     }
where 
\mbox{${\bar \gamma}_{\alpha} 
  = \exp [ {\langle |\gamma_{\alpha}|^2 
                \ln |\gamma_{\alpha}| \rangle}
               /{\langle |\gamma_{\alpha}|^2 \rangle } ]$}. 
The inequality 
\Equation{eq:hpUpperLim}
{
     h_{\rm P} \leq \frac{\Delta_{\alpha 0}}{\sqrt{2}} 
     }
can be proved in general 
as shown in \mbox{Appendix\ref{app:Noteineq}}. 
The equality sign holds 
if and only if $\gamma_{\alpha}({\hat \vk})$ is constant.

The real upper critical field would be smaller than 
the value of $h_{\rm c}$ given by \eq.{eq:hcFFLO} 
owing to negative effects, 
such as the orbital pair-breaking effect and the fluctuation effect. 
However, the value of $h_{\rm c}$ is useful 
because a higher $h_{\rm c}$ must imply 
that the free energy of the FFLO state at 
$h \equiv \ \mu_{\rm e} H \sim \Delta_0$ is lower. 
The value of $h_{\rm c}$ can be regarded as 
an index of the strength of the positive effects that stabilize 
the FFLO state.

{\it Systems and States to be Examined} --- 
In the following, we consider a three-dimensional structure 
of the order parameter 
by assuming 
$$
     \gamma_{\alpha}({\hat \vk}) 
     = \gamma_{\alpha}(\varphi,k_z) 
     = \gamma_{\alpha_{\parallel}}^{\parallel}(\varphi) 
       \gamma_{\alpha_{z}}^{z}(k_z)
     $$ 
with 
$\alpha = (\alpha_{\parallel},\alpha_z)$, 
where $\gamma_{\alpha_{\parallel}}^{\parallel}$ 
and $\gamma_{\alpha_z}^{z}$ are basis functions 
that satisfy 
\Equationnoeqn{eq:normalization} 
{
     \int_0^{2 \pi} \frac{d\varphi}{2 \pi} 
     |\gamma_{\alpha_{\parallel}}^{\parallel}(\varphi)|^2 
     = 
     \int_0^{2 \pi} \frac{d k_z}{2 \pi} 
     |\gamma_{\alpha_{z}}^{z}(k_z)|^2 
     = 1 . 
     }
The Fourier transformation of 
$\langle c_{i \uparrow} c_{j \downarrow} \rangle$ 
verifies that $\gamma_{\alpha_z}^{z}(k_z)$ does not depend on $k_z$ 
if $\langle c_{i \uparrow} c_{j \downarrow} \rangle \ne 0$ 
only for sites $i$ and $j$ on the same layer, 
whereas 
$\gamma_{\alpha_z}^{z}(k_z)$ depends on $k_z$ owing to 
the factor $e^{\pm i \vk \cdot (\vR_i - \vR_j)}$ 
if $\langle c_{i \uparrow} c_{j \downarrow} \rangle \ne 0$ 
for sites $i$ and $j$ on different layers, 
where $c_{i \sigma}$ denotes the operator of the electron 
with spin $\sigma$ on site $i$ at $\vR_i$. 
We call the states of the former and latter cases 
the intralayer and interlayer states, 
which are induced by 
intralayer and interlayer attractive interactions, 
respectively.~\cite{NoteInterl} 
We define the one-particle (electron or hole) energy 
\Equation{eq:xidef}
{ 
     \xi_{\vk} 
       =  \epsilon_{\vk_{\parallel}}^{\parallel}
                 - 2 t_z \cos k_z - \mu , 
     }
where 
\Equation{eq:xiparamxmy}
{
     \epsilon_{\vk_{\parallel}}^{\parallel} 
       =   \frac{k_x^2}{2 m_x} 
         + \frac{k_y^2}{2 m_y} 
     }
with $\vk_{\parallel} = (k_x,k_y)$~\cite{Noterho}. 
Equation~\refeq{eq:xiparamxmy} can be transformed into 
an isotropic dispersion 
$\epsilon_{\vk_{\parallel}}^{\parallel} = {\tilde \vk}_{\parallel}^2/2m$ 
by defining 
${\tilde \vk}_{\parallel} = ({\tilde k}_x,{\tilde k}_y)$, 
${\tilde k}_{\mu} = \sqrt{m/m_{\mu}} k_{\mu}$, 
and $m = \sqrt{m_x m_y}$, 
and hence, 
all the equations for $m_x \ne m_y$ 
can be transformed into 
the corresponding equations for $m_x = m_y = m$ 
(Appendix~\ref{app:NoteSim}). 
Therefore, we use 
\Equation{eq:xipara}
{
     \epsilon_{\vk_{\parallel}}^{\parallel}
       =   \frac{{\vk_{\parallel}}^2}{2 m} 
     } 
in the following for conciseness.

The present model can be effective for systems in which 
the electron (or hole) density is small. 
For an arbitrary $\epsilon_{\vk_\parallel}^{\parallel}$, 
redefining the momentum coordinate appropriately 
and, if necessary, 
making the electron-hole transformation, 
we can assume that the minimum of $\epsilon_{\vk_\parallel}^{\parallel}$ 
is at $\vk_{\parallel} = (0,0)$. 
Expanding $\epsilon_{\vk_\parallel}^{\parallel}$ 
around $\vk_{\parallel} = (0,0)$ 
and 
redefining the $k_x$- and $k_y$-axes along the principal axes, 
we obtain \eq.{eq:xiparamxmy} 
as an approximate form 
when the carrier density is small~\cite{NoteSL}.

\section{Systems with Straight Fermi Surfaces} 
\label{sec:tz0}

Before examining systems with $t_z \ne 0$, 
let us examine systems with $t_z = 0$. 
When $t_z = 0$, 
the Fermi surfaces are straight (elliptic) cylinders. 
The cylindrical system 
has been examined in previous studies 
when the magnetic field is parallel to the layers; 
however, the study in this section covers wider situations, 
including 
an arbitrary direction of $\vH$ except for the c-direction, 
interlayer pairing as well as 
intralayer pairing, 
and systems with effective mass anisotropy 
$m_x \ne m_y$. 
Equation~\refeq{eq:falphageneral} reduces to 
\Equation{eq:falpha_straightFS}
{
     f_{\alpha}(\vq) 
     = 
     - 
       \int_0^{2 \pi} \frac{d \varphi}{2 \pi} 
       |\gamma_{\alpha_{\parallel}}^{\parallel}(\varphi)|^2 
       \ln \bigl | 
         1 - \qbar \cos \varphi \, 
           \bigr | , 
     }
where $\qbar = v_{\rm F} q_{\parallel}/2h_{\rm c}$, 
$v_{\rm F} = k_{\rm F}^{\parallel}/m$, 
$q_{\parallel} = |\vq_{\parallel}|$, 
and $\vq_{\parallel} = (q_x,q_y)$. 
Here, $\gamma_{\alpha_z}^{z}(k_z)$ 
has disappeared from the equation; 
hence, the argument in this section 
does not depend on $\alpha_{z}$.

First, we examine the states in which 
the symmetry of $\Delta_{\vk}$ is s-wave in each layer 
and arbitrary in the $k_z$-direction.~\cite{NoteArb} 
In such states, 
$\gamma_{\rm s}^{\parallel}(\varphi) = 1$, 
and hence, 
$f_{\alpha}(\vq) = f_0(\qbar)$; i.e., 
\Equation{eq:fs2D}
{
     f_{\alpha}(\vq) 
     = 
       - \int_0^{2 \pi} \frac{d \theta}{2 \pi} 
       \ln \bigl | 
          1 - \qbar \cos \theta \, 
           \bigr | . 
     } 
Therefore, we obtain 
\Equation{eq:s-wave}
{
     f_{\alpha}(\qbar) 
     = 
     \left \{ 
     \begin{array}{lc}
     \dps{ 
         - \ln ( \bar q /2 )  } 
           & \mbox{~~for~~} \qbar \ge 1 , \\
     \dps{ 
       - \ln \Bigl [ 
             \bigl \{ {1 + (1 - \qbar^2)^{1/2} 
             \bigr \} /2 
             \Bigr ] } 
           } 
           & \mbox{~~for~~} \qbar \le 1 , 
     \end{array}
     \right . 
     }
which is the same as the equation 
in the previous paper~\cite{Shi94} 
except for the definition of ${\bar q}$ 
and the extended applicability mentioned above. 
From \eq.{eq:hcFFLO}, 
it follows that $h_{\rm c} = \Delta_{\alpha 0}$ and $\qbar = 1$. 
This result holds 
for an arbitrary symmetry in the $k_z$-direction 
and an arbitrary magnetic-field direction 
except for $\vH \parallel \vc$.\cite{NoteSng}

\begin{figure}[htbp]
\begin{center}
\begin{tabular}{cc}
\includegraphics[width=2.2cm]
{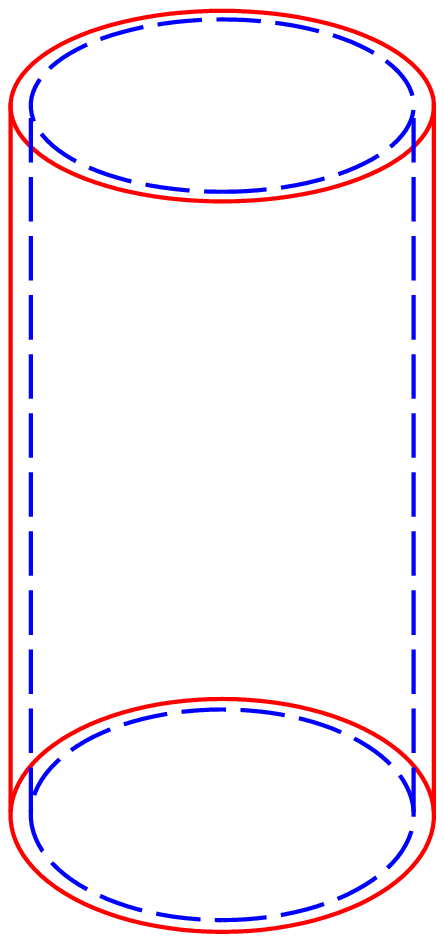}
& 
\includegraphics[width=2.2cm]
{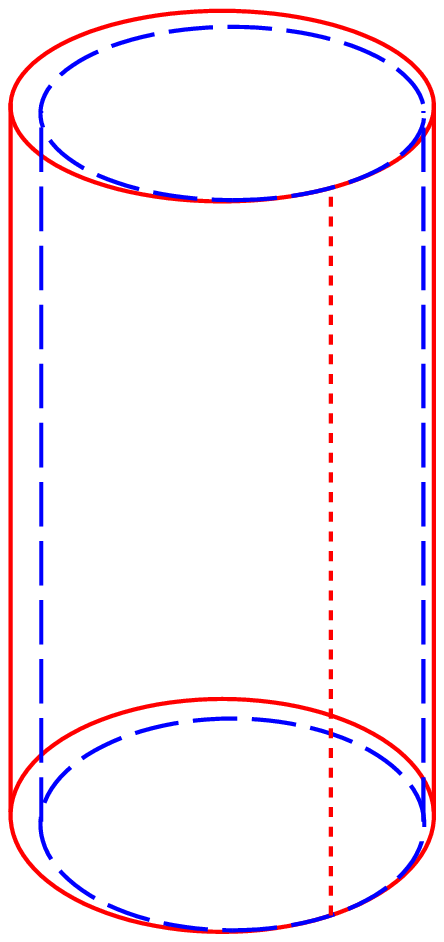}
\\[-8pt]
(a) & (b) 
\\ 
\end{tabular}
\end{center}
\caption{
(Color online) 
Schematics of 
cylindrical Fermi surfaces for $t_z = 0$ 
in the first Brillouin zone. 
(a) 
The red solid and blue dashed curves 
show the Fermi surfaces of the down- and up-spin electrons, 
respectively, 
that split because of the Zeeman energy. 
(b) The Fermi surface of the up-spin electrons is shifted 
by a nesting vector $\vq_0$ that has 
$q_{\parallel} = 2h_{\rm c}/v_{\rm F}$, 
for which the Fermi surfaces touch on the vertical line 
shown by the red dotted line. 
}
\label{fig:FS2D}
\end{figure}

The fact that $H_{\rm c} = \Delta_{\alpha 0}/\mu_{\rm e}$ 
is much larger than 
the Pauli limit 
$H_{\rm P} \leq \Delta_{\alpha 0}/\sqrt{2}\mu_{\rm e}$ 
can be attributed to 
the nesting effect~\cite{Shi94,Shi97a,Shi99aT0,Shi99aFT,Miy14,Ita18}. 
As shown in Fig.~\ref{fig:FS2D}, 
the Fermi surfaces touch on a line (hereinafter called the nesting line) 
by a displacing vector $\vq_0$ that has 
$q_{\parallel} = 2h_{\rm c}/v_{\rm F}$. 
The FFLO vector $\vq$ obtained above 
also has the same $q_{\parallel}$ ($ = 2h_{\rm c}/v_{\rm F}$), 
and the large value of $h_{\rm c}$ can be attributed to 
the fact that the Fermi surfaces touch each other.

Next, we examine the states in which the symmetry of $\Delta_{\vk}$ 
is d-wave in each layer and arbitrary 
in the $k_z$-direction.~\cite{NoteArb} 
For such states, we adopt 
$\gamma_{\rm d}^{\parallel}(\varphi) 
= \sqrt{2} \cos (2 \varphi)$ 
as the principal in-plane basis function. 
As many authors have reported~\cite{Notedw}, 
$h_{\rm c}$ depends on the direction of $\vH_{\parallel}$, 
where $\vH = (\vH_{\parallel},H_z)$. 
Let $\vq_0$ denote the nesting vector that is parallel to $\vH$ 
and has $q_{\parallel} = 2 h_{\rm c}/v_{\rm F}$. 
When $\vH_{\parallel} \parallel \va$, 
because $\Delta_{\vk}$ is maximum on the nesting line, 
the FFLO vector $\vq$ is equal to $\vq_0$, 
as in the s-wave state. 
By contrast, 
when $\vH_{\parallel} \parallel [1,1,0]$, 
because $\Delta_{\vk}$ vanishes on the nesting line, 
we obtain $\vq \ne \vq_0$. 
This is shown by an explicit calculation as outlined below. 
It can be easily verified that 
\Equation{eq:dwfd}
{
     f_{\alpha}(\vq) = 
     \left \{ 
     \begin{array}{ll}
     f_0(\qbar) + f_4(\qbar) 
       & \mbox{~~for~~} \vH_{\parallel} \parallel \va , \\
     f_0(\qbar) - f_4(\qbar) 
       & \mbox{~~for~~} \vH_{\parallel} \parallel [1,1,0] , 
     \end{array} 
     \right . 
     }
and the functions $f_0$ and $f_4$ 
can be obtained as shown in Appendix\ref{app:Notefn}. 
This results in $h_{\rm c}$ and $\qbar$ 
as summarized in Table~\ref{table:dw}. 
The value $\qbar = 1$ for $\vH_{\parallel} \parallel \va$ 
implies that the Fermi surfaces 
touch on a line. 
By contrast, 
the value $\qbar \approx 1.210 > 1$ 
for $\vH_{\parallel} \parallel [1,1,0]$ 
implies that the Fermi surfaces 
are intersected by two vertical lines. 
Because of the nesting effect for the FFLO state, 
$h_{\rm c}$ for $\vH_{\parallel} \parallel \va$ 
is approximately $30\%$ larger than 
that for $\vH_{\parallel} \parallel [1,1,0]$.

\begin{table}[hbtp]
\caption{
Results for the states with a symmetry that is d-wave in each layer 
and arbitrary in the $k_z$-direction, 
when $t_z = 0$ and $\vH \nparallel \vc$. 
}
\label{table:dw}
{\footnotesize 
\begin{center}
\vspace{2ex}
\begin{tabular}{c|cc}
\hline 
 & & \\[-8pt]
  &  $\vH_{\parallel} \parallel \va$  
  &  $\vH_{\parallel} \parallel [1,1,0] $   
  \\[2pt]
\hline 
 & & \\[-6pt]
Max. of $f_{\rm d}$  
  &  $\dps{ \ln 2 + \frac{1}{4} }$  
  &  $\dps{ \frac{1}{2} \ln(\sqrt{3} + 1) + \frac{\sqrt{3} - 1}{4} }$ 
  \\[6pt]
$\qbar$ 
  &    1   
  &    $\sqrt{2} \, (\sqrt{3}-1)^{1/2} \approx 1.210$ 
  \\[2pt] 
Fermi surfaces 
  &    Touch on a line 
  &    Intersected by two lines 
  \\[2pt] 
$\dps{ 
  \frac{h_{\rm c}}{\Delta_{\rm d0}}
  }$   
  &  $e^{1/4} \approx 1.284$  
  &  $\dps{ 
      \frac{1}{2} 
      (\sqrt{3} + 1)^{1/2} 
      e^{\frac{\sqrt{3} - 1}{4}} 
      \approx 0.992
      }$ 
  \\[6pt]
\hline 
\end{tabular}
\end{center}
}
\end{table}

\section{Systems with Warped Fermi Surfaces}\label{sec:finitetz}

In this section, 
we examine systems with $t_z \ne 0$ 
in perpendicular fields, 
i.e., $\vH \parallel \vc$, for which $\vq = (0,0,q)$. 
We assume that $t_z$ is sufficiently small 
so that the Fermi surfaces are open in the $k_z$-direction. 
For the states with $\alpha = (\alpha_{\parallel},\alpha_z)$, 
we obtain 
\Equation{eq:falphaWC}
{
     f_{\alpha}(\vq) 
     = 
     - 
       \int_0^{2 \pi} \frac{d k_z}{2 \pi} 
       |\gamma_{\alpha_z}^{z}(k_z)|^2 
       \ln \bigl | 
         1 - \qbar \sin k_z \, 
           \bigr | , 
     }
where $\qbar$ is redefined as $\qbar = t_z q/h_{\rm c}$. 
Because $\gamma_{\alpha_{\parallel}}^{\parallel}(\varphi)$ 
has disappeared from the equation, 
the following argument does not depend on the in-plane symmetry 
$\alpha_{\parallel}$.

For the intralayer states, 
because \mbox{$\gamma_{\alpha_{z}}^{z}(k_z) = 1$}, 
\eq.{eq:falphaWC} is reduced to 
$f_{\alpha}(\vq) = f_0(\qbar)$, 
i.e., 
\Equation{eq:falphaWC2}
{
     f_{\alpha}(\vq) 
     = 
     - \int_0^{2 \pi} \frac{d k_z}{2 \pi} 
       \ln \bigl | 
         1 - \qbar \sin k_z \, 
           \bigr | , 
     }
which is mathematically equivalent to \eq.{eq:fs2D}; 
however, $\cos \theta$ in \eq.{eq:fs2D} originates from 
the relative angle $\theta$ 
between $\vq_{\parallel}$ and $\vk_{\parallel}$, 
whereas $\sin k_z$ in \eq.{eq:falphaWC2} originates from 
the variation of the Fermi velocity in the $k_z$-direction. 
The same calculation leads to exactly the same equation 
as \eq.{eq:s-wave} 
except for the definition of $\qbar$, 
and hence, we obtain 
$h_{\rm c} = \Delta_{\alpha 0}$ and $\qbar= 1$. 
Interestingly, 
the results coincide for the completely different systems, 
as summarized in Table~\ref{table:s_WC}. 
The fact that $h_{\rm c}$ exceeds the Pauli limit 
$h_{\rm P} = \Delta_{\alpha 0}/\sqrt{2}$ 
implies that the FFLO state can be stabilized, 
which is expected from the results of previous studies.\cite{NoteAdaHou} 
A new finding of the present study is that 
$h_{\rm c}$ {\it substantially} exceeds $h_{\rm P}$ 
and it is of the same order as that for parallel fields. 
This implies that the nesting effect 
significantly enhances the stability of the FFLO state, 
for perpendicular as well as parallel fields.

\begin{table}[hbtp]
\caption{
Comparison of the results for the two types of nesting effect 
for \mbox{$\vH \nparallel \vc$} and \mbox{$\vH \parallel \vc$}. 
}
\label{table:s_WC}
{\footnotesize 
\begin{center}
\vspace{2ex}
\begin{tabular}{c|cc}
\hline 
  & & \\[-8pt]
  Assumptions 
  & & 
  \\[2pt]
\hline 
  & & \\[-8pt]
Interlayer transfer 
  &  $t_z = 0$ 
  &  $t_z \ne 0$ 
  \\[2pt]
Fermi surfaces 
  &  Cylinders 
  &  Warped cylinders 
  \\[2pt]
Direction of $\vH$ 
  &  Arbitrary ($\nparallel \vc$) 
  &  $\vH \parallel \vc$ 
  \\[2pt]
$\gamma_{\alpha_{\parallel}}^{\parallel}$ 
  &  Constant (s-wave) 
  &  Arbitrary 
  \\[2pt] 
$\gamma_{\alpha_{z}}^{z}$ 
  &  Arbitrary 
  &  \begin{tabular}{c}
     Constant \\[-2pt]
     (intralayer pairing)
     \end{tabular}
  \\[2pt] 
\hline 
  & & \\[-8pt]
  Results 
  & & 
  \\[2pt]
\hline 
  & & \\[-8pt]
  Equation 
  & \eq.{eq:fs2D} 
  & \eq.{eq:falphaWC2}
  \\[2pt]
Optimum $\qbar$ 
  &    
       \begin{tabular}{c}
       1 \\
       ($q_{\parallel} = 2 h_{\rm c}/v_{\rm F}$) 
       \end{tabular}
  &    
       \begin{tabular}{c}
       1 \\
       ($q = q_z = h_{\rm c}/t_z$) 
       \end{tabular}
  \\[8pt]
Nesting 
  &    Vertical line 
  &    Horizontal circle 
  \\[0pt]
$h_{\rm c}$ 
  &    $\Delta_{\alpha 0}$ 
  &    $\Delta_{\alpha 0}$ 
  \\[2pt]
\hline 
\end{tabular}
\end{center}
}
\end{table}

\begin{figure}[htbp]
\begin{center}
\begin{tabular}{cc}
\includegraphics[width=2.2cm]
{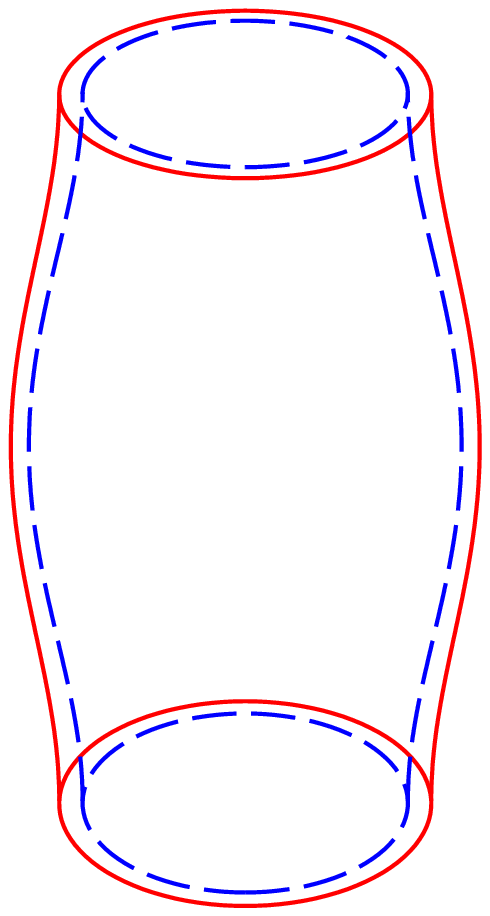}
& 
\includegraphics[width=2.2cm]
{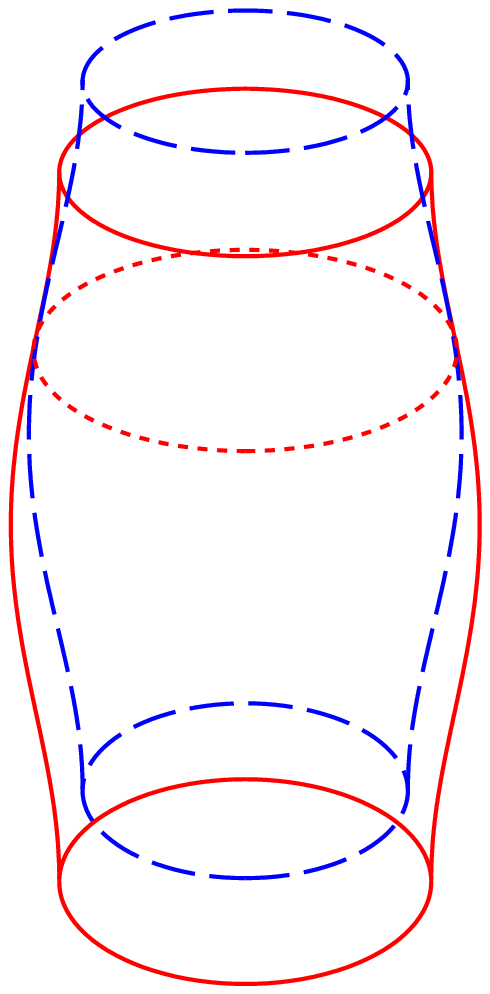}
\\[-8pt]
(a) & (b) 
\\ 
\end{tabular}
\end{center}
\caption{
(Color online) 
Similar to Fig.~\ref{fig:FS2D}, 
but the Fermi surfaces are warped 
and $\vq \parallel \vc$. 
In (b), 
the Fermi surface of up-spin electrons is shifted by 
$\vq_0$ with $|\vq_0| = h_{\rm c}/t_z$. 
For this value of $|\vq_0|$, 
the Fermi surfaces touch on the circle 
shown by the red dotted curve. 
} 
\label{fig:FSWC}
\end{figure}

Although the {\it exact} coincidence in $h_{\rm c}$ 
is only a consequence of the simplifications of the models, 
which lead to the mathematical similarity 
of \eqs.{eq:fs2D} and \refeq{eq:falphaWC2}, 
it is physically significant 
that $h_{\rm c}$ is of the same order in the two cases. 
The physical reason is interpreted using Fig.~\ref{fig:FSWC}. 
The split Fermi surfaces shown in Fig.~\ref{fig:FSWC}(a) 
touch on a circle (the red dotted curve, 
hereinafter called the nesting curve) 
when one of the Fermi surfaces is shifted by $\vq \parallel \vc$ 
as shown in Fig.~\ref{fig:FSWC}(b). 
The nesting curve is a full circle (or ellipse) 
because all the cross sections are circular (or elliptic). 
In general, when the shapes of the cross sections 
perpendicular to the $k_z$-axis are the same, 
the Fermi surfaces touch in a similar manner 
when one of them is shifted by 
the nesting vector $\vq_0$ ($\parallel \vc$) 
appropriate for the Fermi-surface geometry. 
For this mechanism, the nesting curve is not necessarily 
closed like a full circle. 
When parts of the cross sections are similar, 
the nesting effect can work.

The quasi-low-dimensionality plays an essential role 
in the present nesting effect even for perpendicular fields. 
In Fig.~\ref{fig:FSWC}(b), 
the smaller Fermi surface 
is completely inside the larger Fermi surface. 
This implies that 
on the nesting curve, the Fermi surfaces touch but do not cross. 
The open structure of the Fermi surfaces in the $k_z$-direction 
favors this behavior because of the presence of 
the inflection points at $k_z = \pm \pi/2$. 
In contrast, when $t_z$ is large, 
the Fermi surfaces are closed and round 
near the $k_z$-axis, 
and analogously with the spherical system, 
$h_{\rm c}$ is lower than $\Delta_{\alpha 0}$.~\cite{NoteLtz} 
Hence, for the present mechanism of the nesting effect 
in perpendicular fields, 
$t_z$ must be sufficiently small. 
The upper limit of $t_z$ below which 
the Fermi surfaces are open and touch on a curve 
decreases as the carrier density decreases.

{\it Interlayer pairing} --- 
In this part, we consider the interlayer states. 
For the interlayer pairing between electrons on adjacent layers, 
$\Delta_{\vk}$ is proportional to 
$\gamma_{\rm cz}^{z} = \sqrt{2} \cos k_z$ 
or 
$\gamma_{\rm sz}^{z} = \sqrt{2} \sin k_z$, 
where the indices $\alpha_{z} = {\rm cz}$ and ${\rm sz}$ 
are defined. 
When $t_z \ne 0$ and $\vH \parallel \vc$, 
because the nesting curve is at $k_z = \pi/2$ 
as shown in Fig.~\ref{fig:FSWC}(b), 
the nesting effect does not work 
for the states with $\Delta_{\vk} \propto \cos k_z$, 
whereas it significantly enhances $h_{\rm c}$ 
for the states with $\Delta_{\vk} \propto \sin k_z$. 
This behavior is shown in the following.

For the interlayer states, we obtain 
\Equation{eq:interlayer}
{
     f_{\alpha}(\vq) = 
     \left \{ 
     \begin{array}{ll}
     f_0(\qbar) + f_2(\qbar) 
       & \mbox{~~for~~} \alpha_z = {\rm sz} \\
     f_0(\qbar) - f_2(\qbar) 
       & \mbox{~~for~~} \alpha_z = {\rm cz} , 
     \end{array} 
     \right . 
     }
where the functions $f_0$ and $f_2$ are 
given in Appendix\ref{app:Notefn}. 
The results for $h_{\rm c}$ and $\qbar$ are 
summarized in Table~\ref{table:interlayer}. 
Equation~\refeq{eq:interlayer} and Table~\ref{table:interlayer} 
are quite similar to \eq.{eq:dwfd} and Table~\ref{table:dw} 
for the d-wave state in parallel fields. 
The argument that applies the relation between the value of $\qbar$ 
and the nesting for the d-wave state also applies to 
that for the present interlayer states. 
As shown in Table~\ref{table:interlayer}, 
the upper critical field 
$h_{\rm c} = e^{1/2} \Delta_{\alpha 0}$ for the sz-wave states 
is much larger than 
$h_{\rm c} = \Delta_{\alpha 0}/\sqrt{2}$ for the cz-wave states, 
although both values are larger than $h_{\rm P}$ 
because of \eq.{eq:hpUpperLim}. 
Their ratio, $e^{1/2}/(1/\sqrt{2}) \approx 2.33$, 
is much larger than the corresponding ratio of $h_{\rm c}$ 
($1.284/0.992 \approx 1.29$) 
for the d-wave state in parallel fields 
shown in Table~\ref{table:dw}.

To quantitatively estimate the critical fields of the interlayer states, 
the orbital pair-breaking effect must be incorporated. 
The possible difference in the orbital pair-breaking effect 
between the intralayer and interlayer states 
can be an interesting subject to study.

\begin{table}[hbtp]
\caption{
Results for interlayer states 
with an arbitrary in-plane symmetry $\alpha_{\parallel}$ 
when $t_z \ne 0$ and $\vH \parallel \vc$. 
}
\label{table:interlayer}
{\footnotesize 
\begin{center}
\vspace{2ex}
\begin{tabular}{c|cc}
\hline 
 & & \\[-8pt]
$\alpha_{z}$ 
  &  sz 
  &  cz 
  \\[2pt]
\hline 
 & & \\[-6pt]
Nodes 
 & $k_z = 0, \pm \pi$ 
 & $k_z = \pm \pi/2$ 
  \\[4pt]
Max. of $f_{\alpha}$  
  &  $\dps{ \ln 2 + \frac{1}{2} }$  
  &  $\dps{ \frac{1}{2} \ln 2   }$ 
  \\[6pt]
$\qbar$ 
  &    1   
  &    $\sqrt{2} \approx 1.414$ 
  \\[2pt] 
Fermi surfaces 
  &    Touch on a circle 
  &    Intersected by two circles 
  \\[2pt] 
$\dps{ 
  \frac{h_{\rm c}}{\Delta_{\rm \alpha 0}}
  }$   
  &  $e^{1/2} \approx 1.649$  
  &  $\dps{ \frac{1}{\sqrt{2}} 
            \approx 0.707 
      }$ 
  \\[8pt]
\hline 
\end{tabular}
\end{center}
}
\end{table}

{\it Compounds CeCoIn$_5$ and FeSe} --- 
The present mechanism may explain the existence of 
the high-field superconducting phases for $\vH \parallel \vc$ 
in the compounds ${\rm CeCoIn}_5$ and ${\rm FeSe}$, 
which are considered to be the FFLO state. 
At least, as illustrated above, 
the perpendicular direction ($\vH \parallel \vc$) 
is not necessarily 
disadvantageous to the nesting effect for the FFLO state 
in quasi-low-dimensional systems. 
For ${\rm FeSe}$, 
a small carrier density and 
nearly cylindrical Fermi surfaces~\cite{Kas14,Han18,Shib20} 
are compatible with the present model in 
\eqs.{eq:xidef} or \refeq{eq:xiparamxmy}. 
For this compound, 
the specific feature 
$\Delta \sim \epsilon_{\rm F}$~\cite{Kas14,Shib20} 
should be incorporated in future research. 
In ${\rm CeCoIn}_5$, 
a first-principles calculation~\cite{Shis02} 
suggests that 
some of the Fermi surfaces are cylindrical but corrugated. 
The present theory may be applicable to those Fermi surfaces. 
For accurate prediction of the FFLO state, 
extremely accurate information 
on the Fermi-surface structure 
would be required~\cite{Shi97a,Ita18}, 
and 
analysis incorporating realistic shapes of 
Fermi surfaces is a future research direction.

\section{Summary and Conclusion}\label{sec:conclusion}

We examined the FFLO state in 
strongly Pauli-limited
quasi-two-dimensional superconductors 
in magnetic fields perpendicular to the ab-plane, 
focusing on the Fermi-surface nesting effect for the FFLO state; 
the orbital pair-breaking effect is neglected, 
except for the locking of the direction of $\vq$ 
in the direction of $\vH$.
The nesting effect for the FFLO state 
can enhance $h_{\rm c}$ for perpendicular fields 
in systems with cylindrical Fermi surfaces 
warped by $t_z \ne 0$. 
For intralayer states, 
the nesting effect in perpendicular fields is as strong as 
that for the s-wave state 
in parallel fields (Table~\ref{table:s_WC}). 
For the interlayer states, 
the nesting effect in perpendicular fields 
can be more pronounced 
because of the $k_z$-dependence in $\Delta_{\vk}$. 
In fact, it was shown that 
states with $\Delta_{\vk} \propto \sin k_z$ 
exhibit $h_{\rm c}/\Delta_{\alpha 0} \approx 1.649$.

For systems with a Maki parameter that is not sufficiently large, 
the upper critical field must be significantly reduced from 
the values of $h_{\rm c}$, 
which were estimated in the absence of the orbital pair-breaking effect. 
However, $h_{\rm c}$ can be regarded as 
the index of the strength of the nesting effect for enhancing 
the stability of the FFLO state. 
For a fixed strength of orbital pair-breaking effect, 
the larger the value of $h_{\rm c}$, 
the larger the real upper critical field 
in the presence of the orbital effect.

In conclusion, 
the nesting effect for the FFLO state can be significant 
for perpendicular fields as well as for parallel fields. 
For the nesting effect in perpendicular fields, 
the value of $t_z$ must be sufficiently small, 
and in particular, 
quasi-low-dimensional systems with open Fermi surfaces 
favor the present mechanism. 
The upper limit of $t_z$ decreases 
as the carrier density decreases.

{\bf Acknowledgments}~~ 
The author would like to thank \mbox{Y. Matsuda} for useful discussions 
on the compound ${\rm FeSe}$ 
and related works.

\mbox{}

\appendix
\section{
Explicit forms of $f_n(p)$
}
\label{app:Notefn}

The integral in \eq.{eq:fn} can be carried out explicitly. 
The results for $n = 0$ are 
\Equationnoeqn{eq:f0}
{
       f_0(p) 
         = \left \{ 
         \begin{array}{ll}
         \dps{ 
         - \ln |\frac{p}{2}| } 
           & \mbox{~~for~~} p \geq 1 , \\[12pt]
         \dps{ 
         - \ln \frac{1 + \sqrt{1 - p^2}}{2} }
           & \mbox{~~for~~} p \leq 1 , \\
         \end{array}
         \right . 
     }
and those for an integer $n \ne 0$ are 
  \Equationnoeqn{eq:fnresult_smallp}
  {
       f_n(p)
         = \left \{ 
         \begin{array}{ll}
         \dps{ 
         \frac{1}{n} \cos \Bigl ( n \arccos \frac{1}{p} 
                          \Bigr )
         } 
         & \mbox{~~for~~} p \geq 1 , \\[12pt]
         \dps{ 
         \frac{1}{n} 
           \Bigl ( 
             \frac{|p|}{ 1 + \sqrt{1 - p^2} } 
           \Bigr )^n 
         } 
         & \mbox{~~for~~} p \leq 1 . \\
         \end{array}
         \right . 
       }
The functions 
$f_2(p)$ and $f_4(p)$ can be expressed as 
$$
      f_2(p) = - \frac{1}{2} + \frac{1}{p^2} , ~~~~~ 
     f_4(p) = 
       \frac{1}{4} - \frac{2}{p^2} + \frac{2}{p^4} 
     $$ 
for $p \ge 1$.

\appendix
\section{
Proof of \eq.{eq:hpUpperLim}
}\label{app:Noteineq}

For an arbitrary function $g(x)$ 
and an arbitrary average $\langle \cdots \rangle$ 
over an arbitrary variable $x$, it can be proved that 
if $\langle g \rangle = 1$, 
\Equation{eq:aveglng}
{
       \langle g \ln g \rangle \geq 0 , 
       }
where the equality sign holds for $g(x) = 1$. 
Applying \eq.{eq:aveglng} to 
the average defined in \eq.{eq:avedef} 
and 
$g = |\gamma_{\alpha}|^2/\langle |\gamma_{\alpha}|^2 \rangle$, 
we obtain 
\mbox{$\sqrt{\langle |\gamma_{\alpha}|^2 \rangle} \leq 
 {\bar \gamma_{\alpha}}$}, 
which leads to \eq.{eq:hpUpperLim}.

{\it Proof of \eq.{eq:aveglng}} --- 
It is sufficient to prove this for a simple average 
  such as 
  \Equation{eq:simpleave}
  {
       \langle g \rangle 
       = \frac{1}{n} \sum_{k = 1}^{n} g_k 
       }
  with an arbitrary positive integer $n$. 
  In fact, an arbitrary probability function $p(x)$ 
  can be realized by a sufficiently dense distribution 
  of $x_k$ ($k = 1,2,\cdots, n$) 
  on the $x$-axis 
as 
$$
     \int d x p(x) g(x) \approx \frac{1}{n} \sum_{k = 1}^{n} g_k, 
     $$ 
  where $g_k = g(x_k)$. 
  For the average defined by \eq.{eq:simpleave}, 
  the inequality in \eq.{eq:aveglng} is easily proved 
  with mathematical induction as follows. 
  For $n = 2$, defining $x$ 
  with $g_1 = 1 + x$ and $g_2 = 1-x$, 
  $f(x) \equiv \langle g \ln g \rangle$ satisfies 
  $f'(x) \geq 0$ and $f(0) = 0$. Hence, $f(x) \geq 0$. 
  When \eq.{eq:aveglng} is satisfied for $n$, 
  \eq.{eq:aveglng} is satisfied for $n + 1$. 
  In fact, 
  assuming $g_{n + 1} \leq 1$ without loss of generality, 
$$
     {\tilde g}_k \equiv \frac{n g_k}{n + 1 - g_{n+1}}
     $$ 
satisfies 
  \Equationnoeqn{eq:gtildeineq}
  {
       \frac{1}{n} \sum_{k}^{n} 
         {\tilde g}_k \ln {\tilde g}_k \geq 0 
       }
  because of the induction hypothesis. 
  Hence, 
\Equationnoeqn{eq:nplus1}
  {
       \frac{1}{n+1} \sum_{k = 1}^{n+1} g_k \ln g_k 
       \geq F(g_{n+1}) \geq 0 
       }
  with 
$$
     F(x) \equiv (n + 1 - x)\ln \frac{n+1-x}{n} + x \ln x. 
     $$ 
  The last inequality holds because 
  $F(1) = 0$ and $F'(x) \leq 0$ for $x \leq 1$. 
  It is evident that the equality sign holds when $g_k = 1$ 
  for all integers $k$.

\appendix
\section{
Effective Mass Anisotropy 
}
\label{app:NoteSim}

It follows from the definitions 
$$
     {\tilde k}_{\mu} = \sqrt{\frac{m}{m_{\mu}}} k_{\mu}, 
     $$ 
and 
$m = \sqrt{m_x m_y}$ 
that 
$dk_xdk_y = d{\tilde k}_x d{\tilde k}_y$ 
and 
$$
     \epsilon_{\vk_{\parallel}}^{\parallel} = {\tilde k}_{\parallel}^2/2m , 
   $$ 
where ${\tilde k}_{\parallel}$ and ${\tilde \varphi}$ are 
defined by 
\Equationnoeqn{eq:deftilde_kphi}
{
     ({\tilde k}_{x},{\tilde k}_{y}) 
       = ({\tilde k}_{\parallel} \cos {\tilde \varphi},  
          {\tilde k}_{\parallel} \sin {\tilde \varphi}) . 
     }
We also obtain 
$\rho = m/2\pi$ and 
$$
     \frac{d^2 {\hat \vk}}{S_0} 
     = \frac{d {\tilde \varphi}}{2\pi} 
       \frac{d k_z}{2 \pi} . 
     $$ 
The FFLO vector $\vq$ is also transformed as 
${\tilde q}_{\mu} = \sqrt{m/m_{\mu}} q_{\mu}$ 
and 
$$
     \frac{\vv_{\rm F} \cdot \vq}{2 h_{\rm c}} 
  =  \frac{{\tilde v}_{\rm F}^{\parallel} 
           {\tilde q}_{\parallel}}
          {2h_{\rm c}} \cos {\tilde \theta} 
    + \frac{t_z q_z}{2h_{\rm c}} \sin k_z , 
    $$
where 
$({\tilde q}_x ,{\tilde q}_y ) 
     = ({\tilde q}_{\parallel} \cos {\tilde \theta}_{\vq}, 
        {\tilde q}_{\parallel} \sin {\tilde \theta}_{\vq})$, 
and ${\tilde \theta} \equiv {\tilde \varphi} - {\tilde \theta}_{\vq}$. 
Hence, all the equations for $m_x \ne m_y$ 
have exactly the same form as those for $m_x = m_y = m$.

Here, we briefly explain the influence of the above transformation 
on the orbital pair-breaking effect. 
The components of the vector potential $\vA(\vr)$, 
which appears in the gap equation 
and induces the orbital pair-breaking effect, 
are scaled with different coefficients 
that depend on the effective masses, 
whereas the Zeeman term is unaffected. 
Hence, the strength of the orbital pair-breaking effect 
and the Maki parameter depend on the in-plane field direction.~\cite{NoteEMA}




\begin{thebibliography}{99}
\bibitem{Note2D} 
  Here, the term ``two-dimensional models'' 
  does not mean purely two-dimensional models, 
  but those in which weak three-dimensional interactions 
  that stabilize the long-range order are implicitly 
  taken into account 
  by applying mean-field approximations.
\bibitem{Notemultq} 
  Below the upper critical field, 
  FFLO states with 
  more than two $\vq$'s 
  can be stable, 
  especially in quasi-low-dimensional systems.~\cite{Shi98} 
  However, 
  because the second-order transition point can be 
  examined by considering a single $\vq$, 
  we do not consider multiple $\vq$ states 
  in the present study.
\bibitem{NoteqpH}
  When the orbital pair-breaking effect is sufficiently weak 
  ($\alpha > 1.8$), 
  the vortex and the FFLO oscillation coexist.~\cite{Gru66} 
  In the coexistence state, 
  $\vq \parallel \vH$ as examined in Ref.~\citen{Gru66}. 
  However, when the orbital effect is very weak, 
  vortex states with higher Landau-level indices $n$ occur, 
  and the oscillation of the order parameter can have 
  components perpendicular to $\vH$.~\cite{Shi09,Leb08} 
  The oscillation is reduced to the FFLO oscillation 
  of the pure FFLO state in the limit $n \rightarrow \infty$. 
  Hence, when $\alpha \gg 1$, 
  practically speaking, 
  the FFLO vector has components perpendicular 
  to $\vH$. 
  See also Refs.~\citen{Kle00} and \citen{Kle04}.
\bibitem{NoteLO} 
  States with $\Delta(\vr) \propto \cos (\vq\cdot \vr)$ 
  have two FFLO vectors, i.e., $\pm \vq$, and hence, 
  $\pm \vq \parallel \vH$; 
  however, 
  for the same reason as that mentioned in Ref.~\citen{Notemultq}, 
  we consider a single $\vq$.
\bibitem{NoteNesting1} 
  This effect is called the nesting effect, 
  in analogy with the nesting effect for the SDW and CDW.
\bibitem{NoteNesting2} 
  Details of the nesting effect were explained 
  in previous papers, such as 
  Refs.~\citen{Shi94,Shi97a,Shi99aT0}, 
  and~\citen{Shi99aFT,Miy14,Ita18}.
\bibitem{NoteSRO} 
  In ${\rm Sr_2RuO_4}$, 
  the FFLO state may occur at high fields parallel to the ab-plane 
  because the antiparallel-spin pairing is considered 
  to be confirmed.~\cite{Pus19} 
  The horizontal line node observed in Ref.~\citen{Kit18} 
  is compatible with the interlayer pairing.~\cite{Has00,Shi20c}
\bibitem{Notelambda}
  At $T = 0$, $\Delta_{\rm s 0} = \Delta_{\rm s}$ 
  for the s-wave state with $\gamma_{\rm s} = 1$, 
  whereas 
  $\Delta_{\alpha 0} = \Delta_{\alpha}/{\bar \gamma_{\alpha}}$ 
  for the other $\alpha$. 
  For example, see Ref.~\citen{Shi99aT0}. 
  The dimensionless coupling constant $\lambda_{\alpha}$ 
  is expressed as $\lambda_{\alpha} = g_{\alpha} N_{\alpha}(0)$, 
  where 
  $N_{\alpha}(0) = N(0) \langle [\gamma_{\alpha}]^2 \rangle$, 
  and 
  $g_{\alpha}$ denotes the coupling constant of 
  the pairing interaction for $\alpha$-wave pairing, 
  which is limited by the cutoff energy $\omega_{\rm c}$.
\bibitem{Notechi} 
  The internal field is enhanced by a factor ${\tilde \chi}/\chi$, 
  where $\chi$ and ${\tilde \chi}$ 
  are the bare susceptibility and 
  the susceptibility renormalized 
  by the repulsive interactions between electrons, 
  respectively. 
  Hence, the FFLO critical field 
  (in terms of the bare external field) 
  is reduced by the factor $\chi/{\tilde \chi} < 1$. 
  On the other hand, because 
  $N_{\alpha}(0)\Delta_{\alpha 0}^2/2 = {\tilde \chi}H_{\rm P}^2/2$, 
  $H_{\rm P}$ is reduced by the factor 
  $\sqrt{\chi/{\tilde \chi}}$. 
  Consequently, 
  $H_{\rm c}/H_{\rm P}$ is reduced by the factor 
  $\sqrt{\chi/{\tilde \chi}} < 1$.~\cite{Shi94,Tak69} 
  Note that the attractive interactions between electrons 
  that induce the superconductivity are 
  effective interactions in which the retardation effect 
  is incorporated. 
  The bare static interactions between electrons on the Fermi surfaces 
  should be repulsive because of the strong Coulomb repulsion, 
  which results in ${\tilde \chi} > \chi$. 
  Hence, 
  if the real $H_{\rm c}$ is to exceed the real $H_{\rm P}$ 
  in the presence of the internal-field enhancement, 
  the difference $H_{\rm c} - H_{\rm P}$ ($> 0$) 
  in the absence of the internal-field enhancement 
  must be substantial.
\bibitem{Noterho}
  Equation~\refeq{eq:xidef} results in $\rho(\xi,{\hat \vk}) = m/2\pi$ 
  and $\vv_{\rm F} = (\vk_{\parallel}/m, 2 t_z \sin k_z)$, 
  and hence, 
  $$
        \frac{\vv_{\rm F} \cdot \vq}{2 h_{\rm c}} 
        =     \frac{v_{\rm F}^{\parallel} q_{\parallel}}
                   {2h_{\rm c}} \cos \theta 
            + \frac{t_z q_z}{2h_{\rm c}} \sin k_z , 
       $$
  where 
  \mbox{$v_{\rm F}^{\parallel} \equiv |\vk_{\parallel}|/m$}, 
  and 
  $\theta$ denotes the angle between $\vk_{\parallel}$ 
  and \mbox{$\vq_{\parallel} = (q_x,q_y)$}.
\bibitem{NoteArb} 
  If we consider the FFLO state in singlet superconductors, 
  the symmetry in the $k_z$-direction should be even and odd 
  when the in-plane state is even and odd, respectively. 
  However, the FFLO state can occur in triplet superconductors 
  as long as the pairing occurs between electrons 
  with antiparallel spins.
\bibitem{NoteSng} 
  The singular behavior at $\vH \parallel \vc$ is an artifact 
  of the mean-field gap equation applied to the present ideal model with $t_z = 0$, 
  and it must be removed by incorporating the reality, 
  such as the fact that $t_z$ should be finite 
  even when it can be omitted in the gap equation\cite{Note2D} 
  and 
  $1/q_z$ cannot be smaller than the spacing between the layers.
\bibitem{NoteLtz} 
  When the Fermi surfaces are closed and round near the $k_z$-axis, 
  and we consider a nesting vector $\vq$ parallel to $\vc$, 
  they can touch only at a point on the $k_z$-axis 
  but cannot touch on a curve in most cases. 
  However, the touch at a point yields a lower $h_{\rm c}$ than 
  that by intersection by curves. 
  Hence, $\vq$ becomes longer so that 
  the Fermi surfaces are intersected by a curve. 
  The resulting $h_{\rm c}$ is lower than the value 
  $h_{\rm c} = \Delta_{\alpha 0}$ given by the touch on a curve 
  in the system with open Fermi surfaces.
\end{thebibliography}
\end{document}